\documentclass[10pt]{article}
\usepackage{amsmath}
\usepackage{graphicx}
\usepackage{lineno}

\def\Title#1{\begin{center} {\Large #1 } \end{center}}
\def\Author#1{\begin{center}{ \sc #1} \end{center}}
\def\Address#1{\begin{center}{ \it #1} \end{center}}

\newcommand\pubblock{\rightline{\begin{tabular}{l} Proceedings of the Fifth Annual LHCP\\ \pubnumber\\
         \pubdate  \end{tabular}}}

\newenvironment{Abstract}{\begin{quotation} \begin{center} 
             \large ABSTRACT \end{center}\bigskip 
      \begin{center}\begin{large}}{\end{large}\end{center} \end{quotation}}

\newenvironment{Presented}{\begin{quotation} \begin{center} 
             PRESENTED AT\end{center}\bigskip 
      \begin{center}\begin{large}}{\end{large}\end{center} \end{quotation}}

\def\beq{\begin{equation}}
\def\eeq#1{\label{#1}\end{equation}}
\def\eeqn{\end{equation}}

\def\beqa{\begin{eqnarray}}
\def\eeqa#1{\label{#1}\end{eqnarray}}
\def\eeqan{\end{eqnarray}}

\let\bar=\overbar

\def\Dslash{\not{\hbox{\kern-4pt $D$}}}
\def\dslash{\not{\hbox{\kern-2pt $\del$}}}

\def\msb{{\bar{\ssstyle M \kern -1pt S}}}

\usepackage{color}
\usepackage{subfigure}
\usepackage{hyperref}
\usepackage{multicol}

\newcommand\pubnumber{ ATL-PHYS-PROC-2017-123 }

\newcommand\pubdate{\today}

\def\affiliation{
On behalf of the ATLAS and CMS Collaborations, \\
CEA Saclay \\
91191 Gif-sur-Yvette Cedex, FRANCE}

\begin{document}

\large
\begin{titlepage}
\pubblock

\vfill
\Title{Lepton and photon performance at ATLAS and CMS}
\vfill

\Author{Arthur Alexis Jules Lesage}
\Address{\affiliation}
\vfill
\begin{Abstract}
These proceedings report on lepton and photon performance results obtained using data recorded in 2015 and 2016 at the LHC by the two experiments ATLAS and CMS. For each particle (electrons, photons and muons), the reconstruction and identification efficiencies are presented for the two experiments together with the isolation studies. Results concerning the electron and photon energy calibration as well as the muon momentum scale and resolution are also reported. Despite more challenging pile-up conditions with respect to Run 1, the two experiments achieved impressive performance in early Run 2.
\end{Abstract}
\vfill

\begin{Presented}
The Fifth Annual Conference\\
 on Large Hadron Collider Physics \\
Shanghai Jiao Tong University, Shanghai, China\\ 
May 15-20, 2017
\end{Presented}
\vfill
\end{titlepage}
\def\thefootnote{\fnsymbol{footnote}}
\setcounter{footnote}{0}
%

\normalsize 


\section{The ATLAS and CMS detectors}

The ATLAS and CMS experiments are composed of a tracker close to the interaction point in the centre of the detector, which provides precise coordinate measurements of the tracks and triggering. Calorimeters surround the tracker and provide measurement of the energy of the particles. The outermost part of the detector is composed of the muon system only reached by muons. On top of detecting the muon tracks, this system improves the measurement of the muon transverse momentum and provides triggering.

The main difference between the two detectors is the magnet system.  Solenoid magnets surround the Inner Detector for ATLAS~\cite{ATLAS_experiment}, whereas they surround the Tracker and calorimeters for CMS~\cite{CMS_experiment}. ATLAS has a 2\,T field, where it reaches 3.8\,T for CMS. This field bends the charged particles to determine their charges and their transverse momenta. In ATLAS toroid magnets are part of the Muon Spectrometer and enable improved measurements of the momentum, whereas CMS uses the central field for the measurements.

\section{Electron performance}

\subsection{Reconstruction and identification efficiencies}

Electron reconstruction relies on information from the tracker and the electromagnetic calorimeter.

In ATLAS, electrons are reconstructed by matching a track from the Inner Detector with clusters of energy in the calorimeters. These clusters are formed using a sliding window algorithm with fixed size. In order to suppress contamination by converted photons, the hits in the innermost layer of the Insertable B-Layer are checked for: no pairs of tracks leading to an invariant mass equal to $0$ are retained. To select good quality electrons, criteria are applied to the cluster and track variables, as well as to a likelihood discriminant built using the distribution of the TRT high-threshold hits. These criteria are gathered into so-called working points (WPs)~\cite{ATLAS_electron_reco_ID_eff,CMS_electron_photon} targeting specific signal efficiencies or background rejections.

The procedure is quite similar for CMS, except that the clusters in the electromagnetic calorimeter are composed of several clusters in the azimuthal-angle direction $\phi$. These are referred to as \emph{superclusters} and overcome the important spread of the energy deposits due to the very strong magnetic field. Converted photons are also vetoed. Identification WPs are defined as a combination of cuts on specific variables. These discriminants are also gathered in a multivariate analysis, whose training is carried out in bins of the transverse momentum $p_{\text{T}}$ and pseudo-rapidity $\eta$.

\begin{figure}
	\centering
	\subfigure[ATLAS, efficiencies of the \emph{Loose}, \emph{Medium} and \emph{Tight} WPs]{\includegraphics[height = 0.22\textheight]{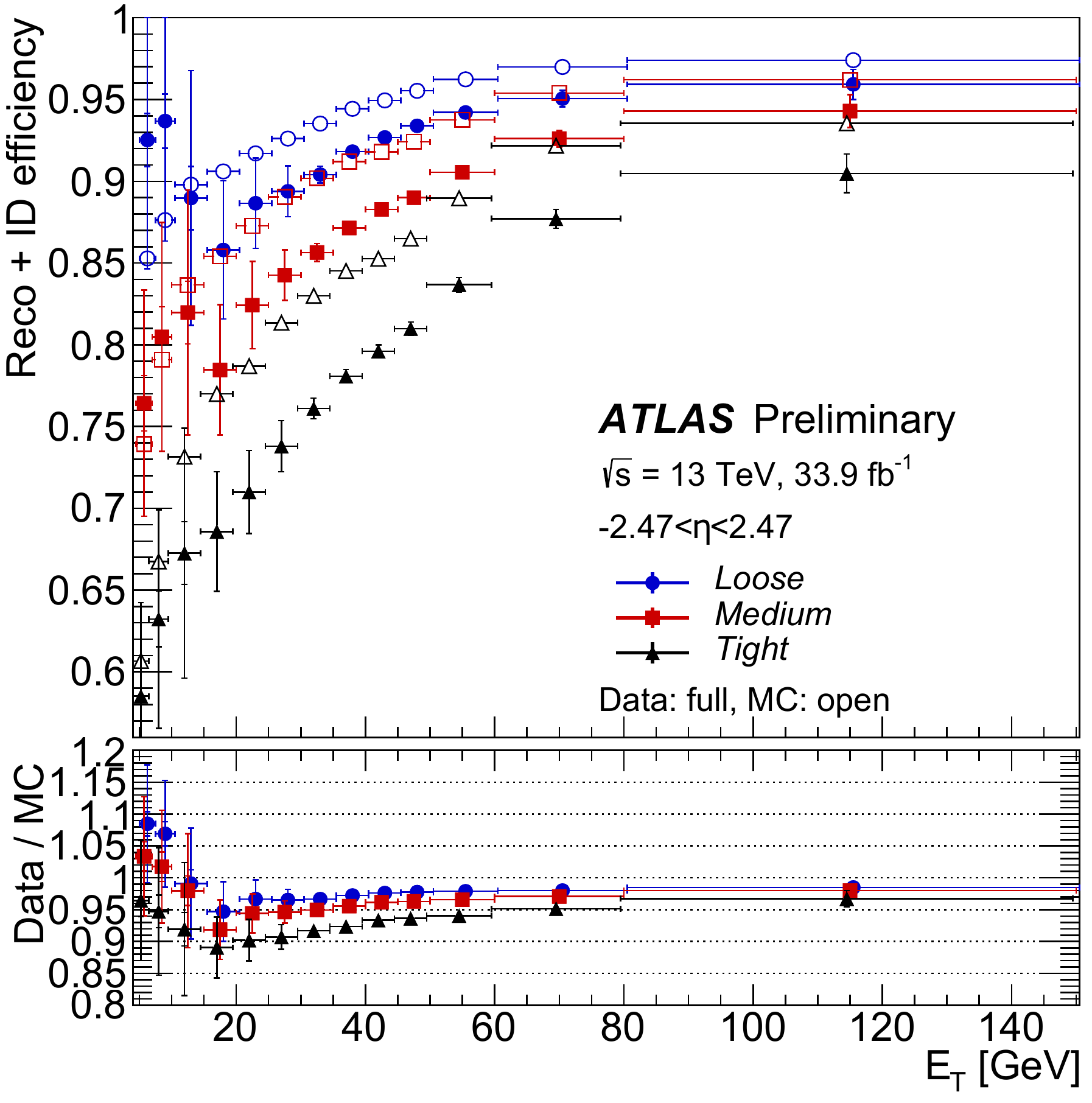}}
	\hspace{0.5 cm}
	\subfigure[CMS, efficiencies of the \emph{Loose} WP in various regions of $\left|\eta\right|$ (including isolation selection)]{\includegraphics[height = 0.22\textheight]{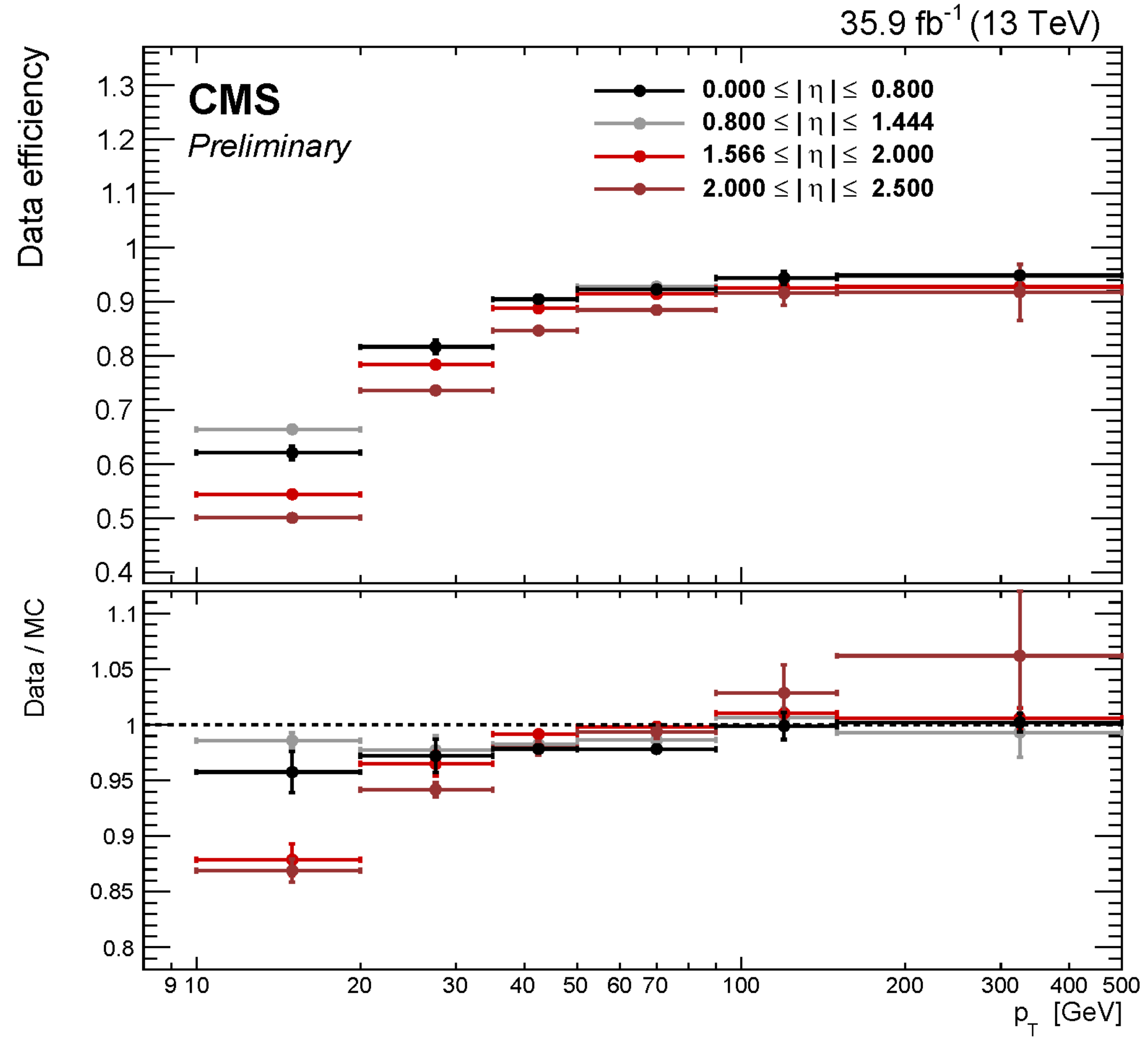}}
	\caption{Electron identification efficiencies measured in $Z\rightarrow{}ee$ and $J/\psi\rightarrow{}ee$ events against $E_{\text{T}}$~\cite{ATLAS_electron_reco_ID_eff,CMS_electron_photon}.\label{fig:electron_reco_ID_eff}}
\end{figure}

Electron reconstruction efficiencies are measured using the tag-and-probe method~\cite{tag_and_probe} on $Z\rightarrow{}ee$ and $J/\psi\rightarrow{}ee$ events. For ATLAS, the \emph{Tight}, \emph{Medium} and \emph{Loose} identification WPs target various signal efficiencies, such that electrons passing the cuts of a WP pass the cuts of a looser one, in the given order. These efficiencies are shown against the transverse energy $E_{\text{T}}$ in Fig.~\ref{fig:electron_reco_ID_eff} (a)~\cite{ATLAS_electron_reco_ID_eff}. The trends are similar with efficiencies above $80\%$ in the barrel (central part of the detector) to $70\%$ in the end-caps. The same results are presented for CMS with the \emph{Tight} and \emph{Loose} identification WPs for various $\eta$ bins, presented against $p_{\text{T}}$ in Fig.~\ref{fig:electron_reco_ID_eff} (b)~\cite{CMS_electron_photon}. Efficiencies are above $70\%$ for $p_{\text{T}} > 20\,\text{GeV}$, up to $90\%$ at higher $p_{\text{T}}$ for the \emph{Loose} WP.

\subsection{Isolation studies}

Isolation consists in assessing the activity surrounding the trajectory of the particle in the tracker and the calorimeters. It is one of the most powerful tools to discriminate signal against background. To calculate the isolation variables, a cone in the $\eta$ and $\phi$ plane is defined around the particle and the energy of close-by objects falling in this cone is added, having subtracted the contribution of the particle itself.

\begin{figure}
	\centering
	\subfigure[ATLAS, efficiency of the \emph{FixedCutLoose} WP against $E_{\text{T}}$]{\includegraphics[height = 0.22\textheight]{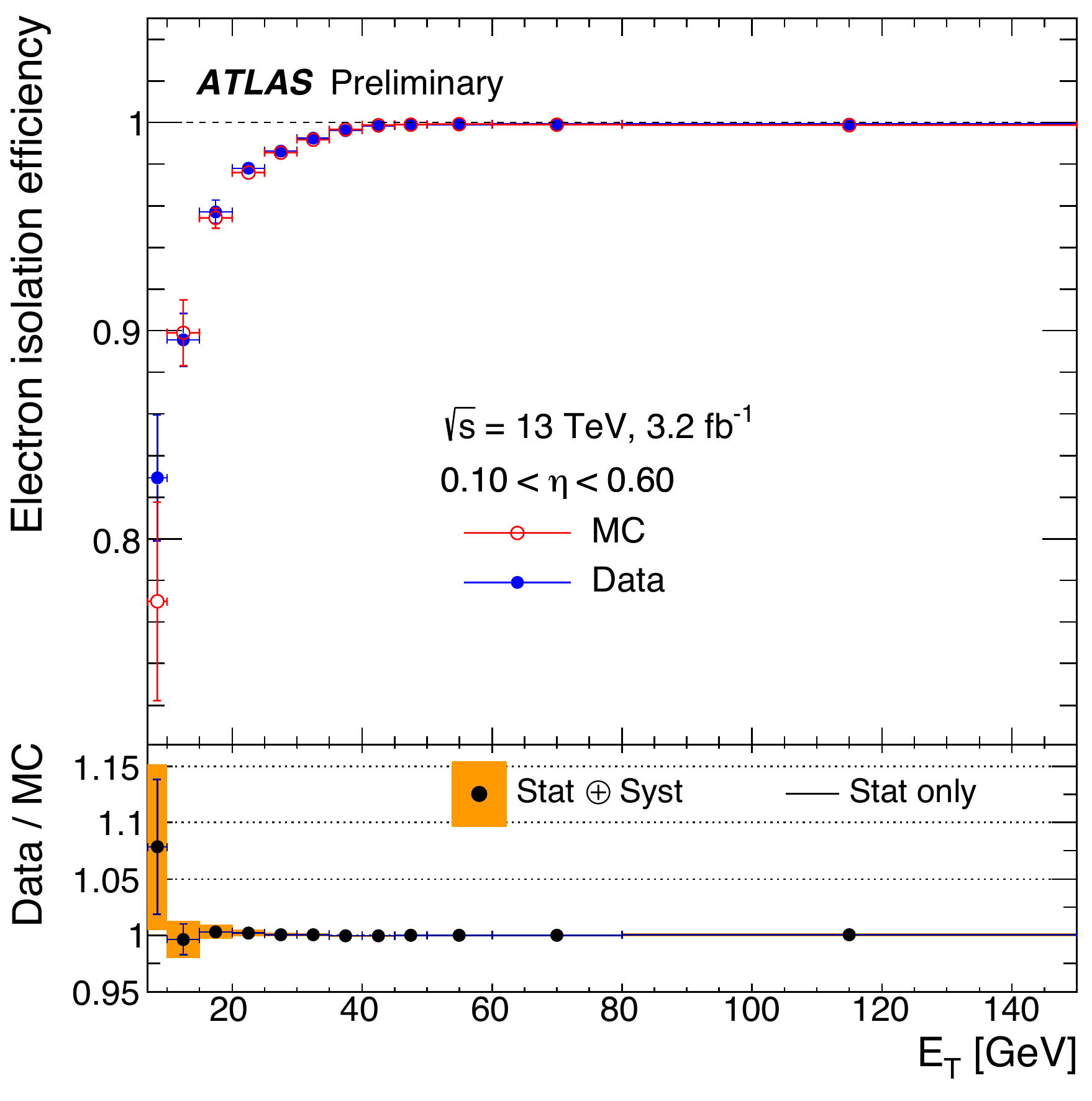}}
	\hspace{0.5 cm}
	\subfigure[CMS, data / simulation comparison of the charged-hadron isolation]{\includegraphics[height = 0.22\textheight]{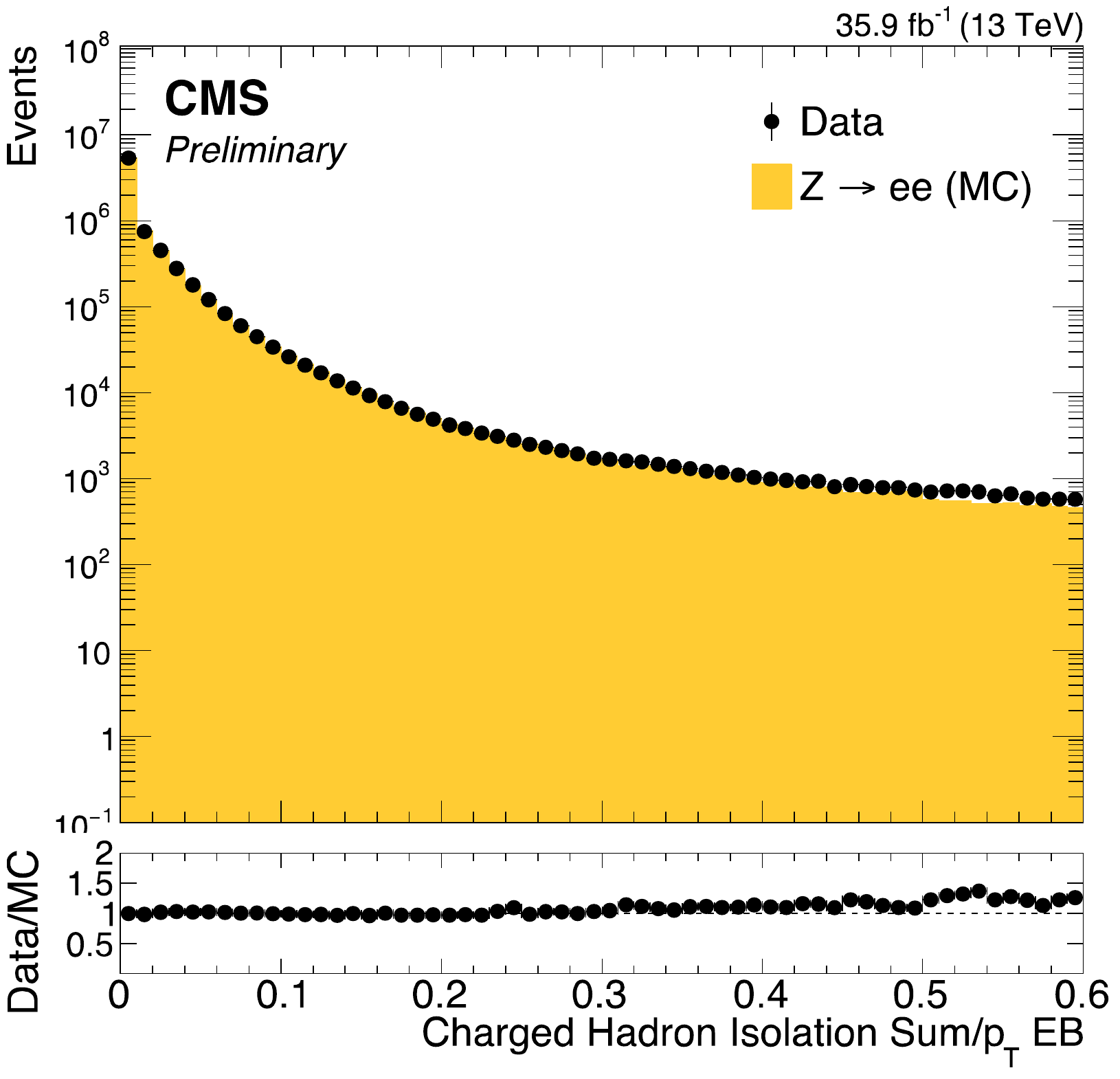}}
	\caption{Electron isolation studies carried out on $Z\rightarrow{}ee$ events~\cite{CMS_electron_photon,ATLAS_electron_iso_eff}.\label{fig:electron_iso}}
\end{figure}

Two kinds of variables are defined in ATLAS. The track-based isolation estimates the surrounding activity by summing up the transverse momenta of the close-by tracks. The calorimeter-based isolation takes instead the energy of clusters of cells close to the particle in the calorimeters. WPs are then defined consisting of cuts applied on the two isolation variables. These cut values are fixed or vary to target specific signal efficiencies as a function of $E_{\text{T}}$. The efficiencies of the \emph{FixedCutLoose} WP against the transverse energy $E_{\text{T}}$ are presented in Fig.~\ref{fig:electron_iso} (a)~\cite{ATLAS_electron_iso_eff}. 

For CMS, discrimination is performed on three kinds of isolation, depending on the nature of the contributions considered: charged hadron, neutral hadron and neutral electromagnetic candidates. The energy of the candidates is assessed using the \emph{particle-flow}~\cite{particle_flow} algorithm, which takes into account information from the relevant detectors to reconstruct the close-by objects and their properties. Unlike ATLAS, the isolation cuts are included in the identification WPs for the electrons and the photons. The distribution of the charged-hadron variable is presented in Fig.~\ref{fig:electron_iso} (b)~\cite{CMS_electron_photon} for electrons in the barrel (in the central part of the electromagnetic calorimeter).

\section{Photon performance}

\subsection{Reconstruction and identification efficiencies}

Photon reconstruction is performed a very similar way as for the electrons, except that no tracks are expected in the tracker, unless the photon has converted to two electrons, in which case the corresponding tracks have to be retrieved. For ATLAS, identification WPs use information from the hadronic and electromagnetic calorimeters. For CMS, the identification WPs are very much like the electron ones, except that cuts and criteria differ. Probability for photons to convert while traversing the tracker material is high. Therefore, a conversion track finding is performed starting from clusters in the electromagnetic calorimeter, to match the tracks in the tracker.

\begin{figure}
	\centering
	\subfigure[ATLAS, efficiencies of photons having $\left|\eta\right| < 1.37$ or $1.52 < \left|\eta\right| < 2.37$]{\includegraphics[height = 0.22\textheight]{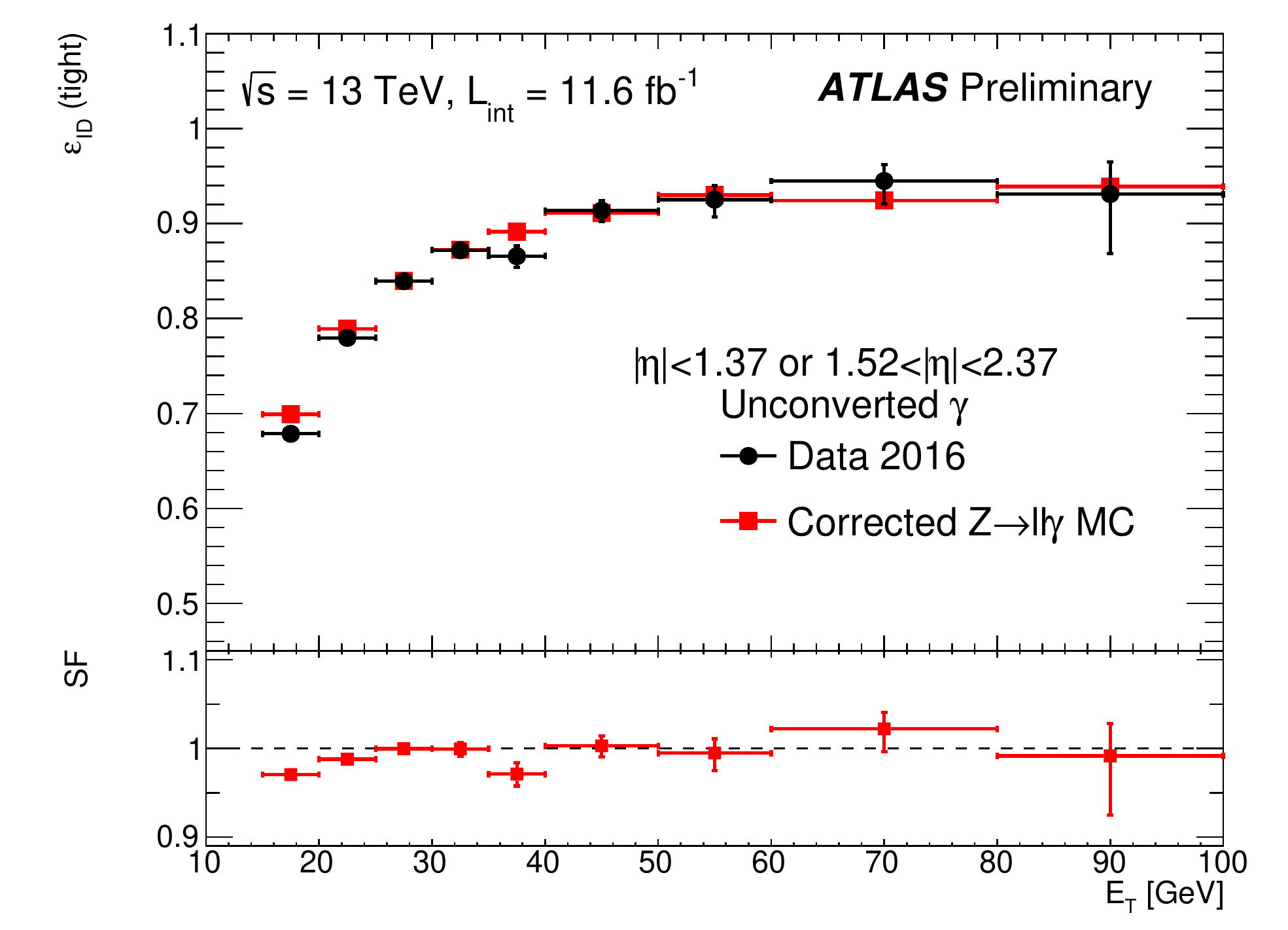}}
	\hspace{0.5 cm}
	\subfigure[CMS, efficiencies given in various $\left|\eta\right|$ regions (including isolation selection)]{\includegraphics[height = 0.22\textheight]{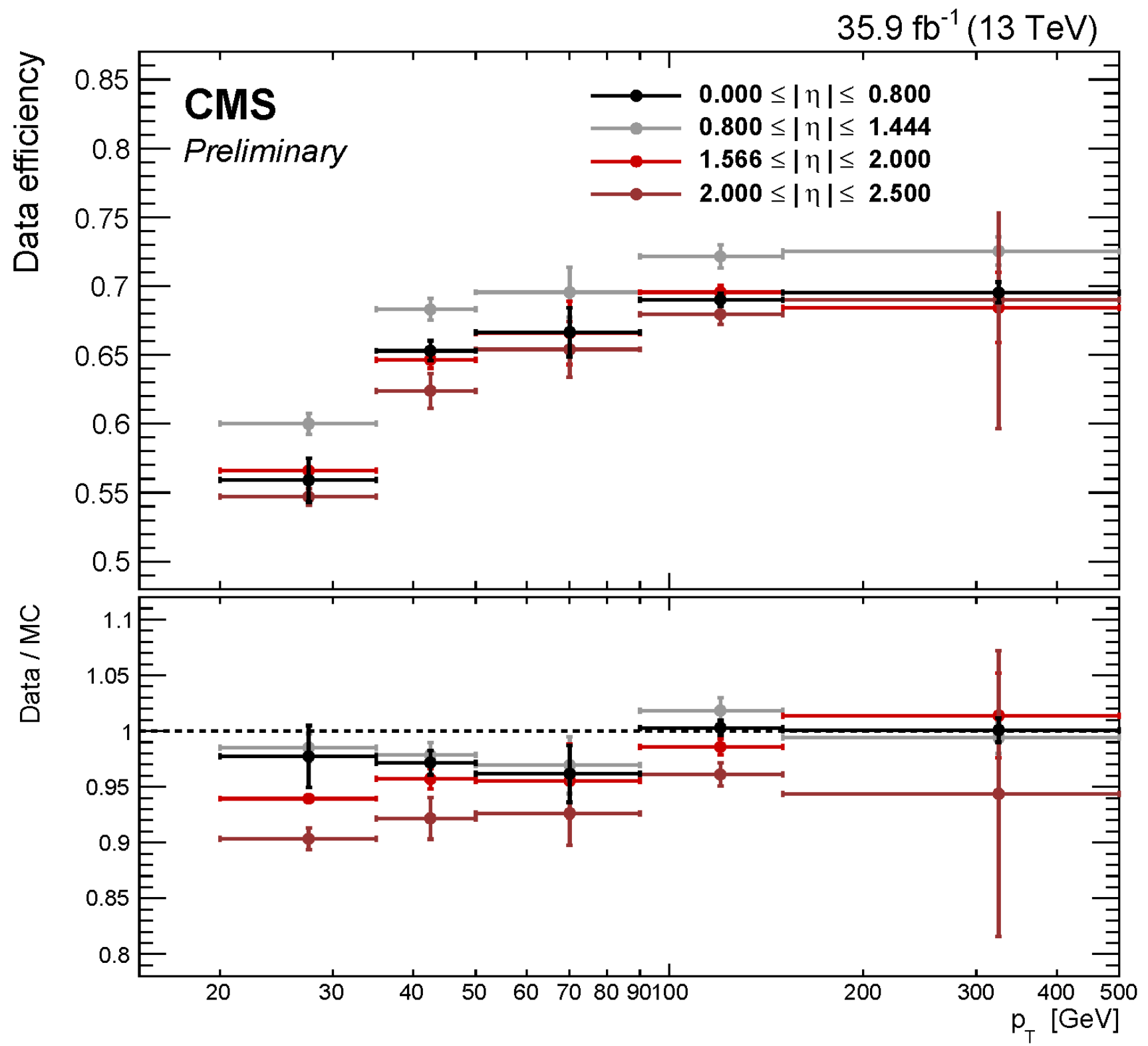}}
	\caption{Photon reconstruction efficiencies of the \emph{Tight} WP against $E_{\text{T}}$~\cite{CMS_electron_photon,ATLAS_photon_reco_ID_eff}.\label{fig:photon_reco_ID_eff}}
\end{figure}

Identification efficiencies are presented for unconverted photons in ATLAS in Fig.~\ref{fig:photon_reco_ID_eff} (a)~\cite{ATLAS_photon_reco_ID_eff}, against $E_{\text{T}}$ for the \emph{Tight} WP. There is a good agreement between data and simulation, and the efficiencies increase with $E_{\text{T}}$.

Similar measurements are performed for CMS, where results are presented in various $\eta$ bins for all photons, as shown in Fig.~\ref{fig:photon_reco_ID_eff} (b)~\cite{CMS_electron_photon}. The \emph{Loose} and \emph{Tight} WPs have efficiencies above $80\%$ and $50\%$ for $p_{\text{T}} > 20\,\text{GeV}$.

\subsection{Isolation studies}

\begin{figure}
	\centering
	\subfigure[ATLAS, \emph{FixedCutTight} efficiency for photons having $\left|\eta\right| < 1.37$ or $1.52 < \left|\eta\right| < 2.37$]{\includegraphics[height = 0.22\textheight]{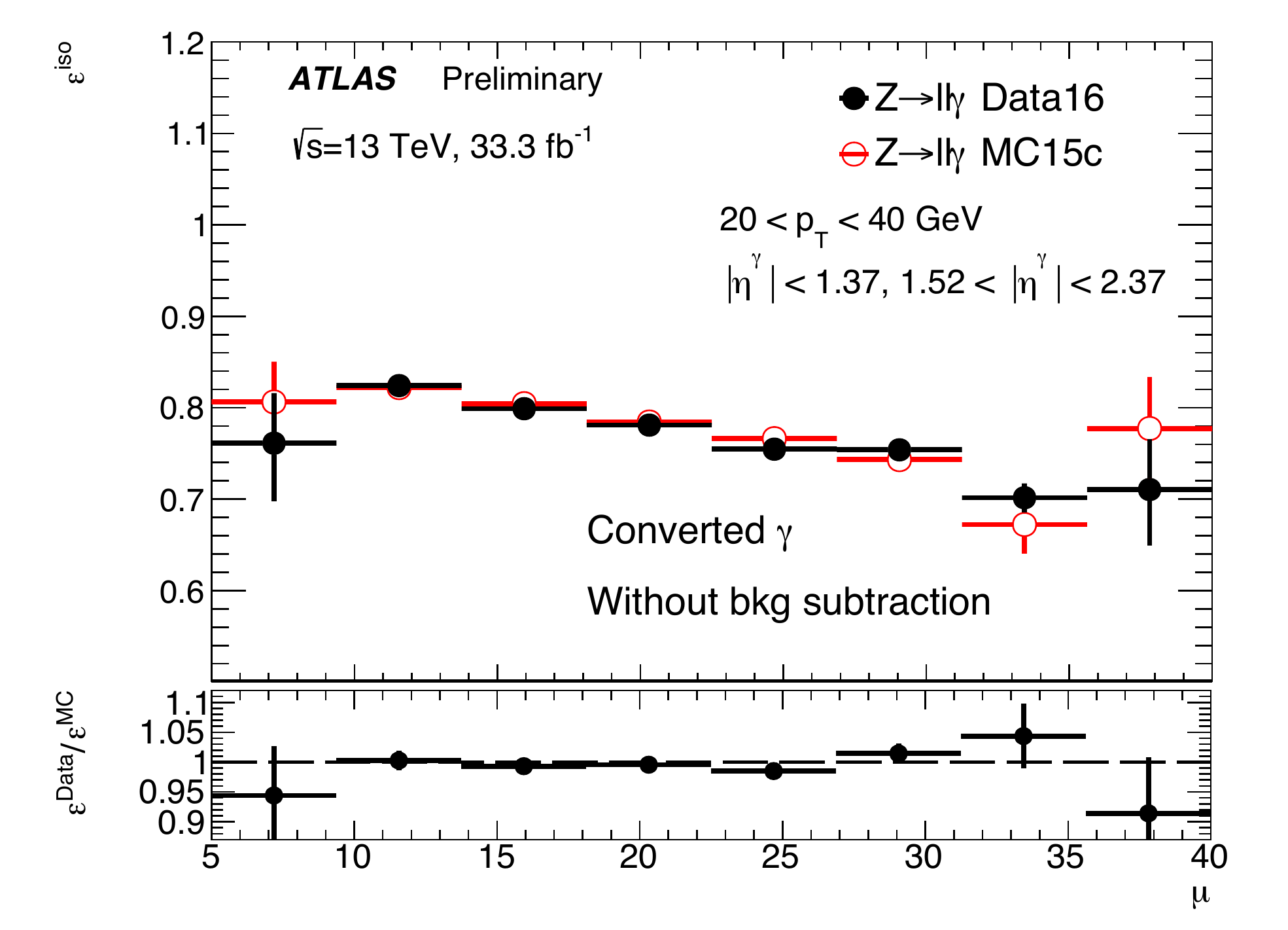}}
	\hspace{0.5 cm}
	\subfigure[CMS, data / simulation comparison of the charged-hadron isolation for barrel photons]{\includegraphics[height = 0.22\textheight]{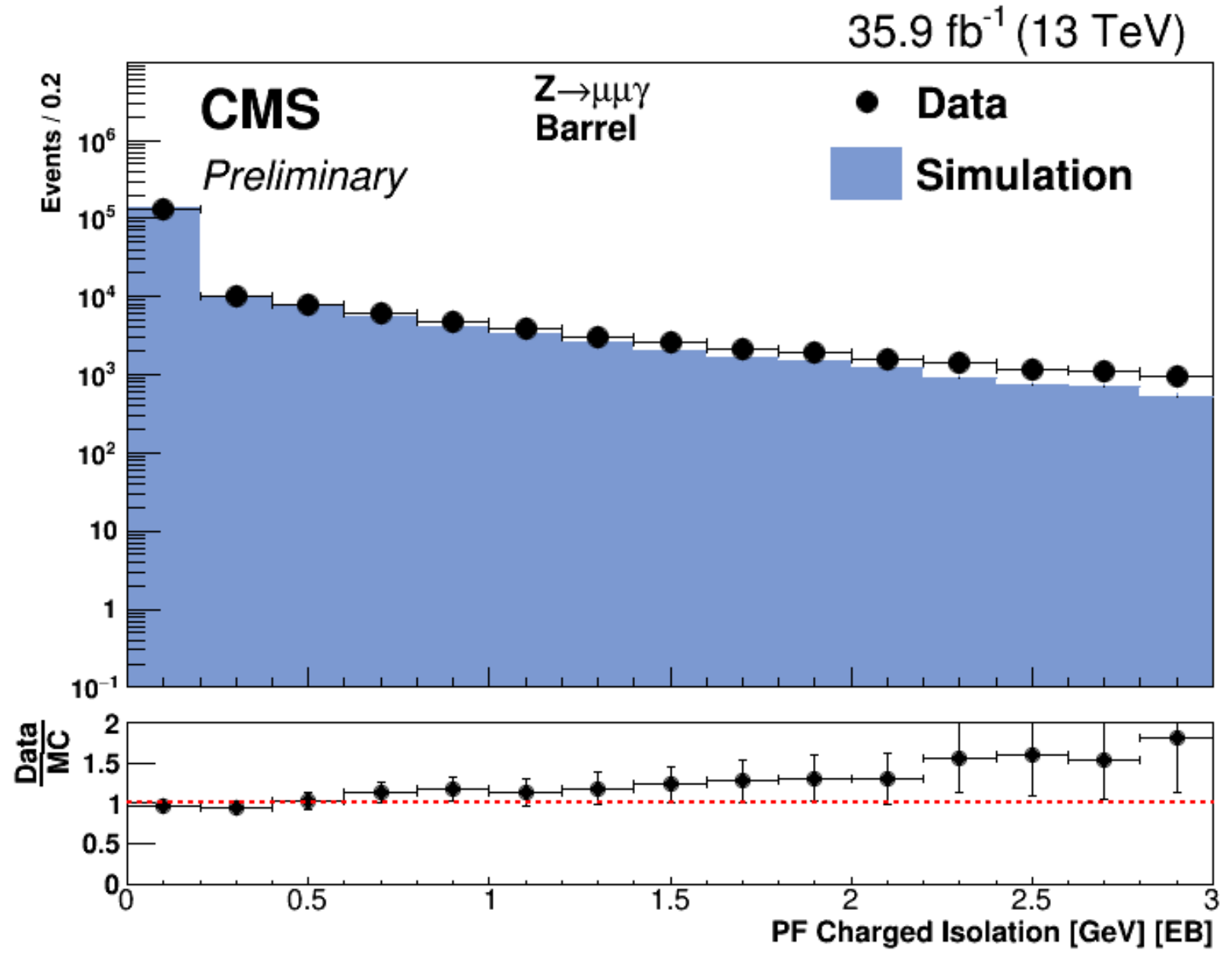}}
	\caption{Photon isolation studies using $Z\rightarrow{}\ell\ell\gamma, \ell\in\left\{e, \mu\right\}$ (ATLAS) and $Z\rightarrow{}\mu\mu\gamma$ (CMS) events~\cite{CMS_electron_photon,ATLAS_photon_iso_eff}.\label{fig:photon_iso_eff}}
\end{figure}

In the ATLAS experiment, the two variables introduced previously are used and cuts values are applied to them to define the \emph{FixedCutTight} WP. The associated signal efficiencies are presented in Fig.~\ref{fig:photon_iso_eff} (a)~\cite{ATLAS_photon_iso_eff}, against the pile-up conditions ($\mu$, the average number of interactions per bunch crossing) for unconverted photons. Thanks to a pile-up correction of the variables, the efficiencies remain robust.

For CMS, the definition of the isolation is similar as for the electron case with isolation variables defined by summing up the contributions of the close-by charged hadrons, neutral hadrons and photons. The results are shown for photons in the barrel in Fig.~\ref{fig:photon_iso_eff} (b)~\cite{CMS_electron_photon}. As for the electrons, isolation cuts are included in the identification WPs.

\section{Electron and photon energy calibration}

Calibration of the electron and photon energy is necessary to correct data and simulation to account for detector response, material description and second-order effects.

In ATLAS, calibration is performed following several steps. A simulation-based calibration is applied to data and Monte Carlo (MC) using a BDT (Boosted Decision Tree algorithm) with gradient boosting. Then, data-driven corrections are optimised to mitigate the non-uniformity in detector response and are applied to data only. Data-driven corrections with energy scale factors 
are applied to data. 
The simulation is then corrected to remove the residual difference with data 
in the energy resolution. 

In CMS, the calibration is more crucial due to the design of the electromagnetic calorimeter, which is composed of crystals whose transparency decreases over time with radiation. The absolute and relative calibrations are calculated among the channels. Corresponding corrections are applied. The energy is finally corrected to account for energy containment effects. This correction is separately tuned for electrons and photons.

\begin{figure}
	\centering
	\subfigure[ATLAS (all electrons)]{\includegraphics[height = 0.22\textheight]{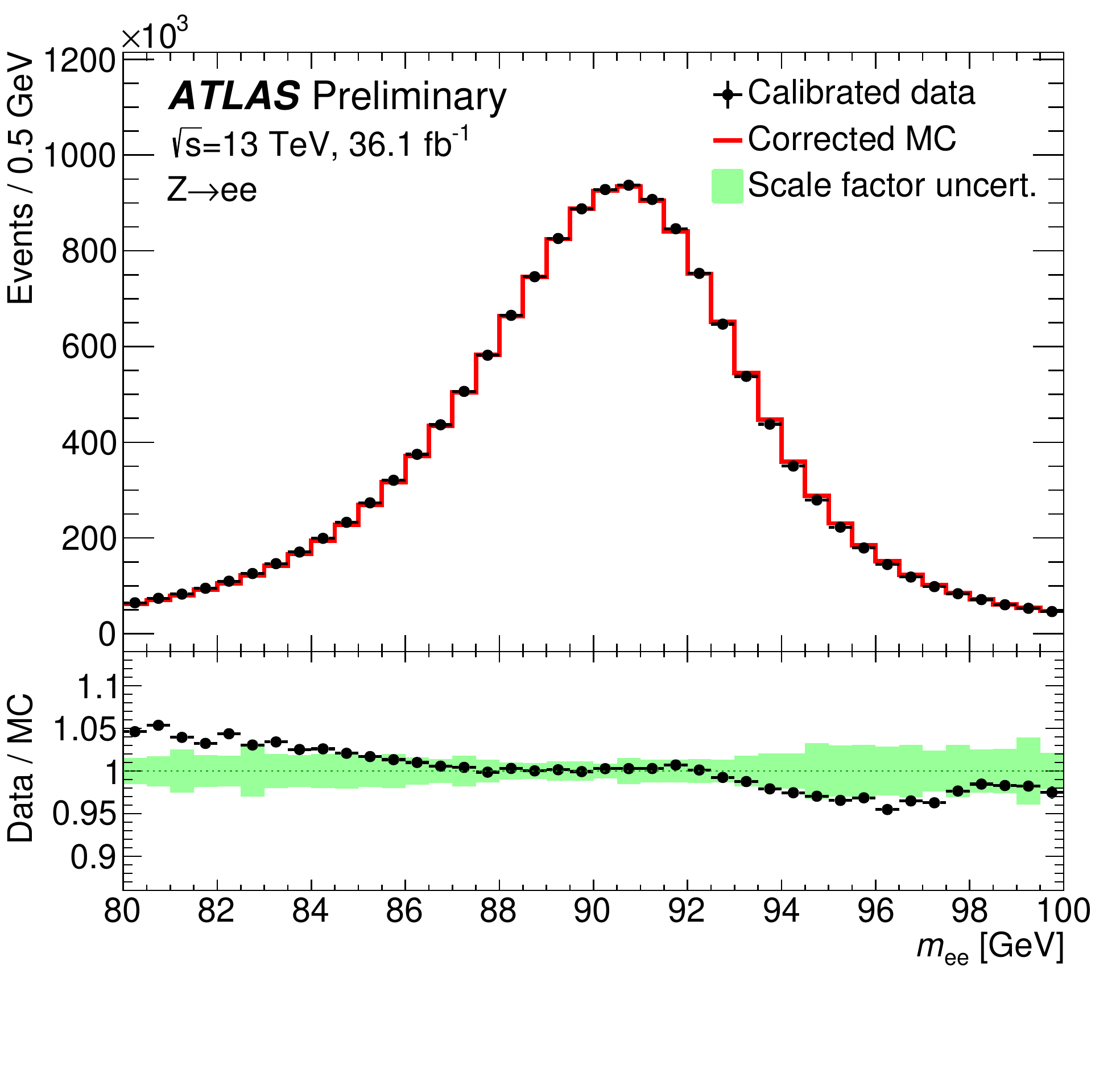}}
	\hspace{0.5 cm}
	\subfigure[CMS (barrel electrons)]{\includegraphics[height = 0.22\textheight]{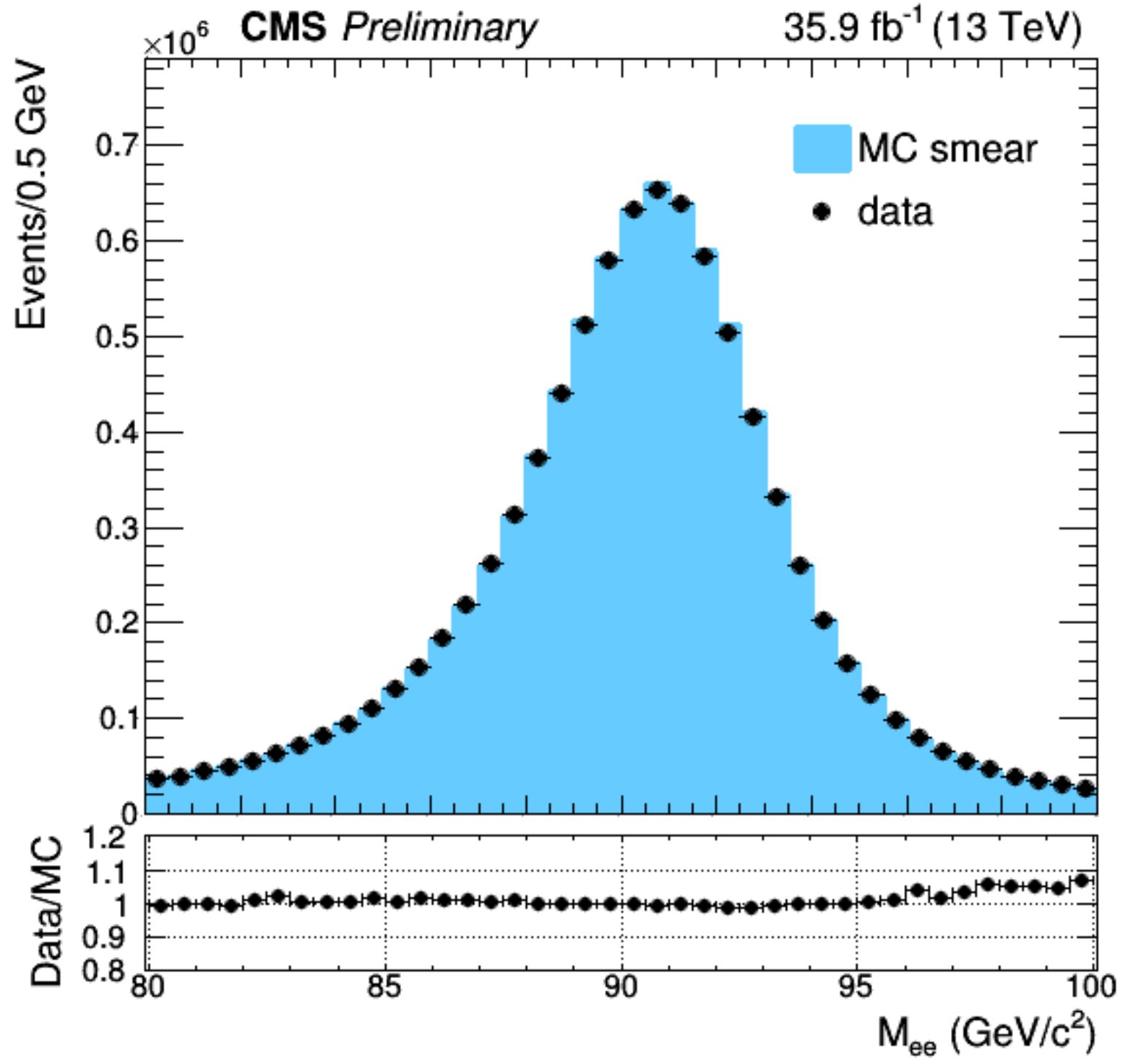}}
	\caption{Comparison data / simulation of the invariant mass distribution in $Z\rightarrow{}ee$ decays after energy calibration~\cite{CMS_electron_photon,ATLAS_electron_photon_energy_calib_res}.\label{fig:electron_photon_energy_calib_res}}
\end{figure}

Results of the calibration are presented using $Z$-boson decays to two electrons, where data are compared to simulation in Fig.~\ref{fig:electron_photon_energy_calib_res}~\cite{CMS_electron_photon,ATLAS_electron_photon_energy_calib_res} for ATLAS and CMS. The two experiments show similar results in terms of resolution which is at the per-cent level in the barrel.

\section{Muon performance}

\subsection{Reconstruction and identification efficiencies}

Muon reconstruction uses information from all the sub-detectors. 

In ATLAS, about $96\%$ of the muons are combined between tracks in the Inner Detector and the Muon Spectrometer. Other muons can be reconstructed tagging Inner Detector tracks or segments in the Muon Spectrometer, with signatures in the calorimeters. This allows the acceptance to increase. Various identification WPs are defined, including criteria on the muon type (combined or not), the quality of the track fit, and the charge over momentum matching between the Inner Detector and the Muon Spectrometer. 

The reconstruction procedure is similar for CMS. Track fits can start from the tracker to the muon system (inside out) or the opposite. This defines the \emph{Tracker} and \emph{Global} muons. \emph{Tracker} Muon reconstruction is more efficient at low momenta. A third type of muon is based on the \emph{particle-flow} algorithm and takes information from all sub-detectors. 
Identification WPs are defined similarly including the same discriminants as ATLAS.

\begin{figure}
	\centering
	\subfigure[ATLAS, \emph{Tight}, \emph{Medium} and \emph{Loose} WPs]{\includegraphics[height = 0.22\textheight]{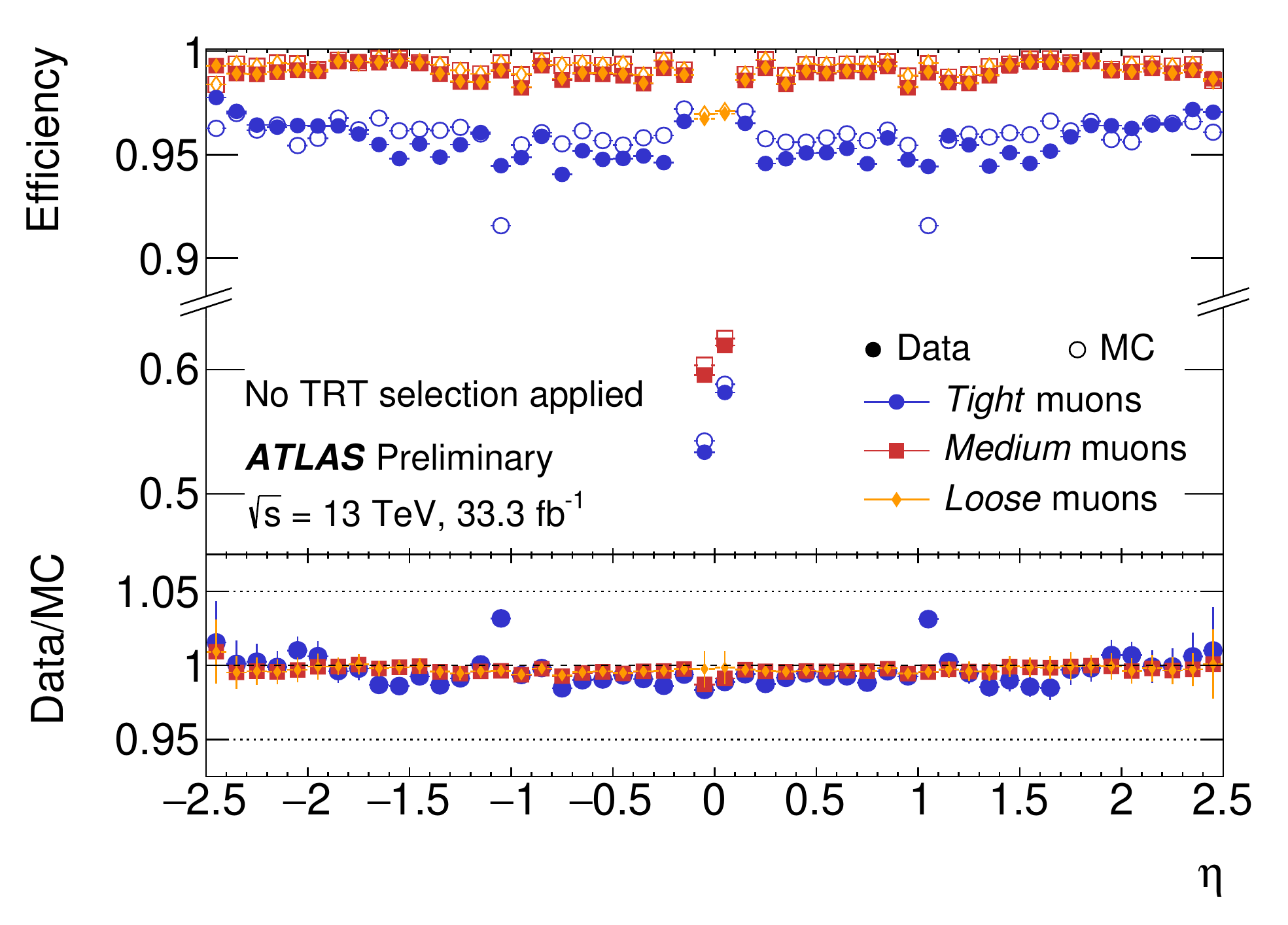}}
	\hspace{0.5 cm}
	\subfigure[CMS, \emph{Loose} WP]{\includegraphics[height = 0.22\textheight]{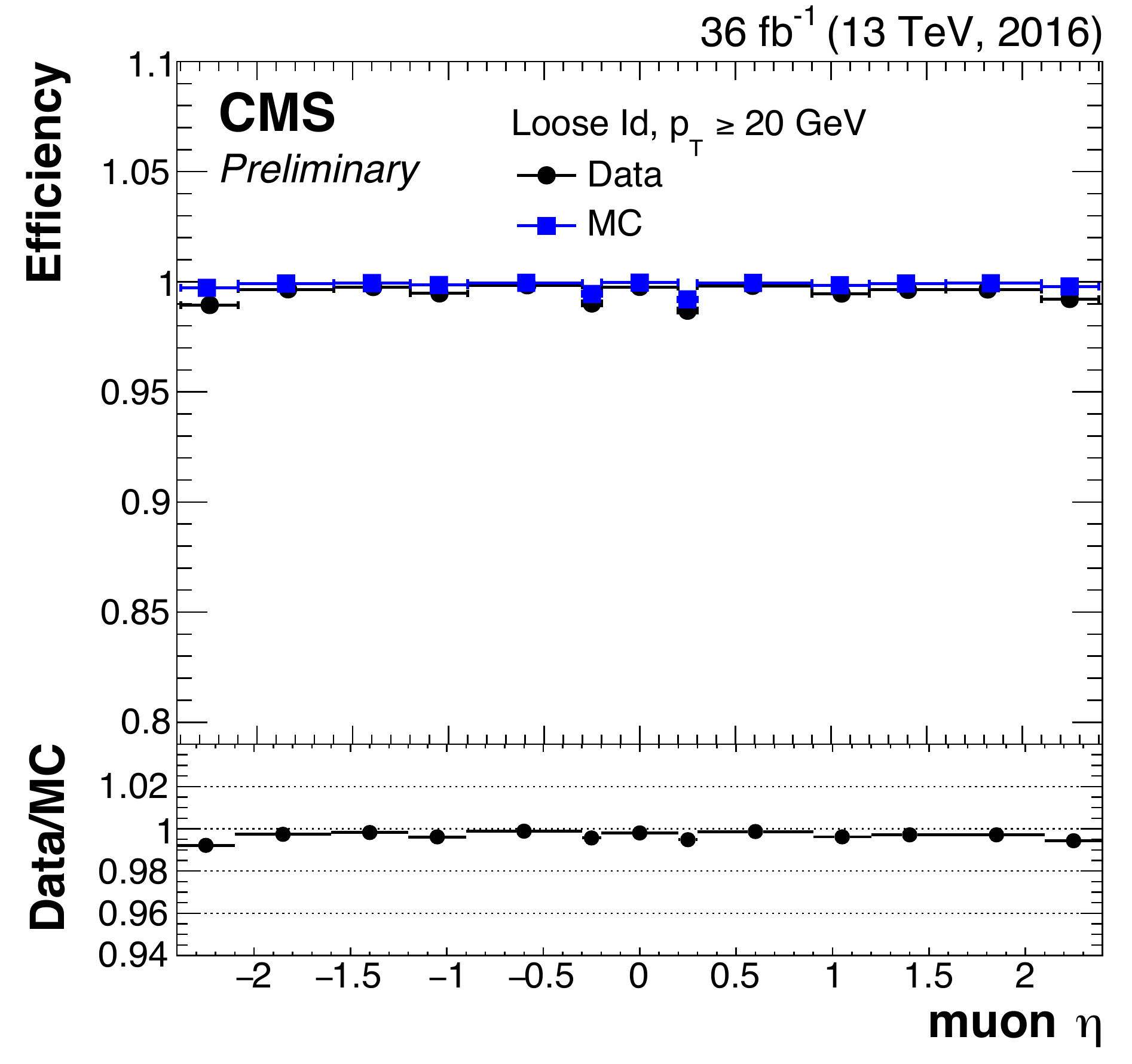}}
	\caption{Muon reconstruction efficiencies measured in $Z\rightarrow{}\mu\mu$ decays, as a function of $\eta$~\cite{ATLAS_muon,CMS_muon}.\label{fig:muon_reco_ID_eff}}
\end{figure}

As for the electrons, the reconstruction efficiencies are calculated using the tag-and-probe method on $Z\rightarrow\mu\mu$ and $J/\psi\rightarrow\mu\mu$ events. For ATLAS, the results are shown in Fig.~\ref{fig:muon_reco_ID_eff} (a)~\cite{ATLAS_muon} using $Z\rightarrow\mu\mu$ decays. Efficiencies are overall above $95\%$, and increase in the very central region for the \emph{Loose} WP as compared to tighter identification WPs. This is due to an acceptance loss of the Muon Spectrometer for cabling in this region, which is recovered by the use of non-combined muons. For CMS, the results of the \emph{Loose} WP are presented in Fig.~\ref{fig:muon_reco_ID_eff} (b)~\cite{CMS_muon} as a function of $\eta$. Efficiencies are above $99\%$ and $95\%$ for muons having $p_{\text{T}} > 20\,\text{GeV}$, similar to ATLAS results.

\subsection{Isolation studies}

\begin{figure}
	\centering
	\subfigure[ATLAS, \emph{FixedCutLoose} WP]{\includegraphics[height = 0.22\textheight]{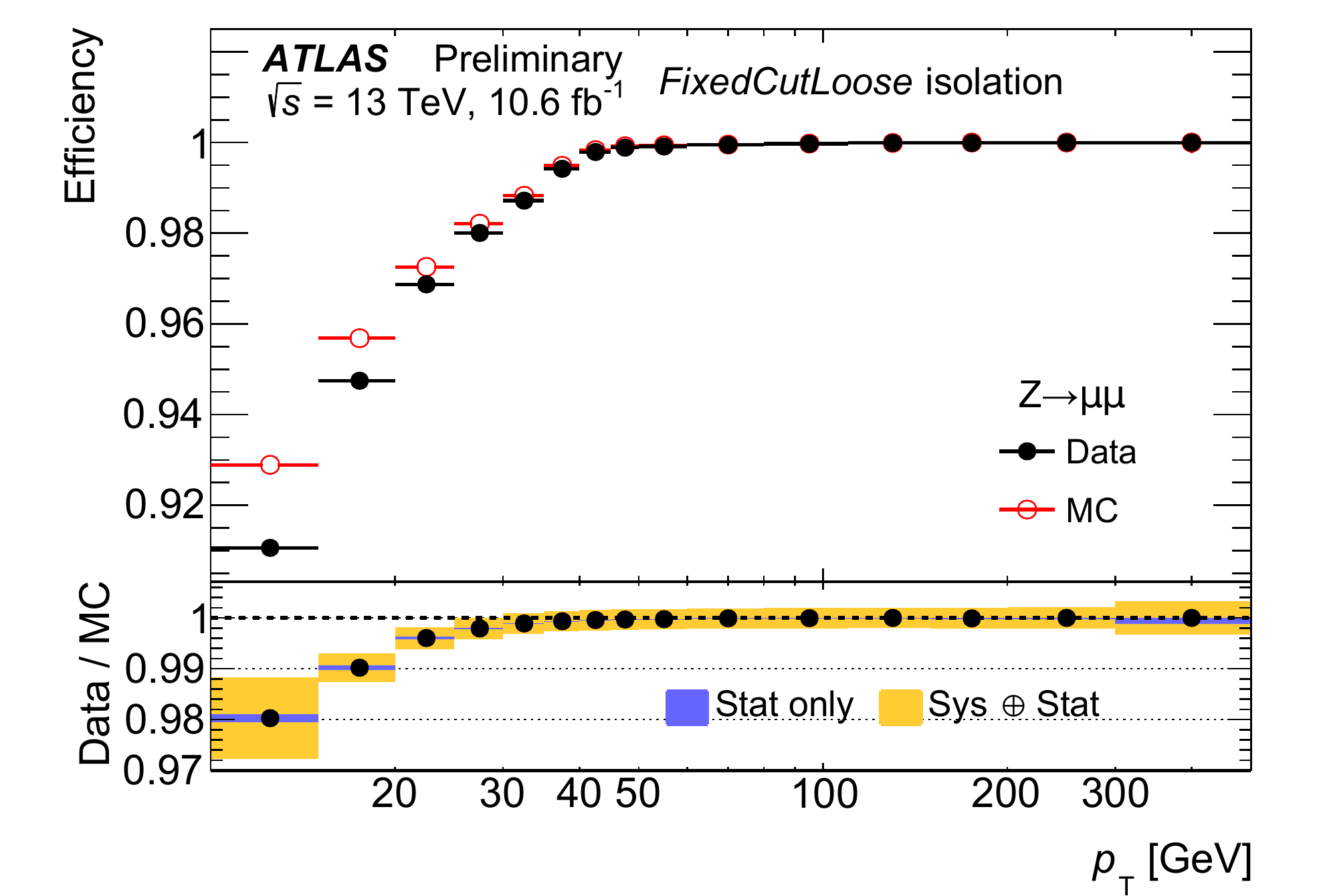}}
	\hspace{0.5 cm}
	\subfigure[CMS, \emph{LooseTracker} WP]{\includegraphics[height = 0.22\textheight]{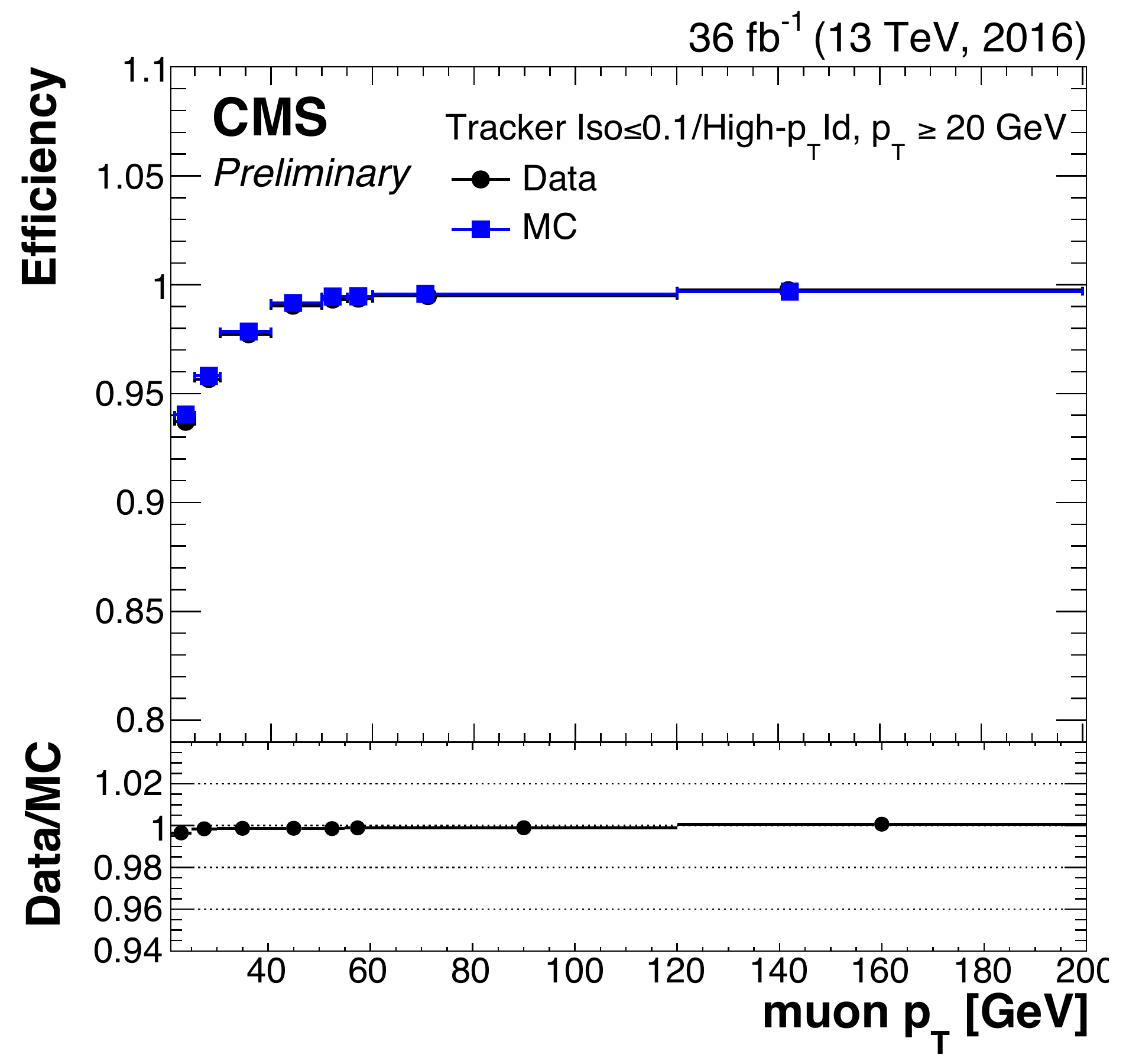}}
	\caption{Muon isolation efficiencies measured in $Z\rightarrow\mu\mu$ events, against $p_{\text{T}}$~\cite{CMS_muon,ATLAS_muon_iso_eff}.\label{fig:muon_iso_eff}}
\end{figure}

Isolation studies in ATLAS are performed the same way as for electrons and photons. The \emph{FixedCutLoose} WP uses fixed cuts on the two variables. The corresponding efficiencies are presented in Fig.~\ref{fig:muon_iso_eff} (a)~\cite{ATLAS_muon_iso_eff}. Agreement between data and MC is good, as shown by the scale factors.

The efficiencies of the \emph{LooseTracker} WP, defined as cuts on the tracker-based isolation, are presented in Fig.~\ref{fig:muon_iso_eff} (b)~\cite{CMS_muon} for CMS. Efficiencies are above $93\%$ and there is a good agreement between data and simulation. Unlike electrons and photons, identification WPs do not include isolation cuts.

\subsection{Momentum scale and resolution}

The simulated muon momentum is corrected to account for imperfect detector modelling and alignment.
  
In ATLAS, the momentum is corrected performing a fit of the $Z$ and $J/\psi$ mass spectra. Corrections are calculated separately for tracks in the Inner Detector and the Muon Spectrometer and are combined afterwards. In CMS, the momentum bias is fitted by linear and constant terms. The $Z$ and $J/\psi$ mass spectra are fit for low and intermediate $p_{\text{T}}$, and cosmic rays are used at high $p_{\text{T}}$. 

\begin{figure}
	\centering
	\subfigure[Dimuon invariant mass scale]{\includegraphics[width = 0.4\textwidth]{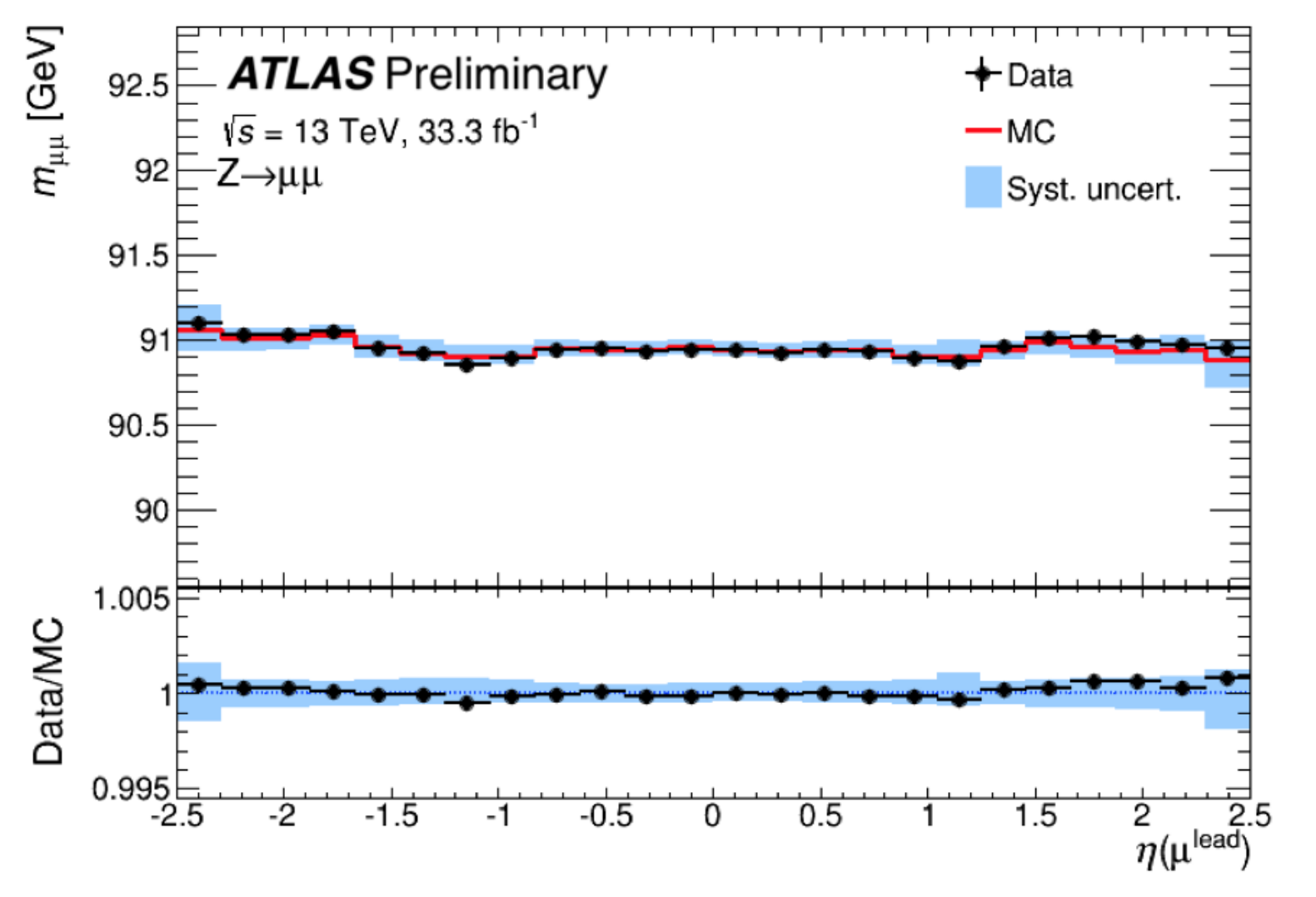}}
	\hspace{0.5 cm}
	\subfigure[Dimuon invariant mass resolution]{\includegraphics[width = 0.4\textwidth]{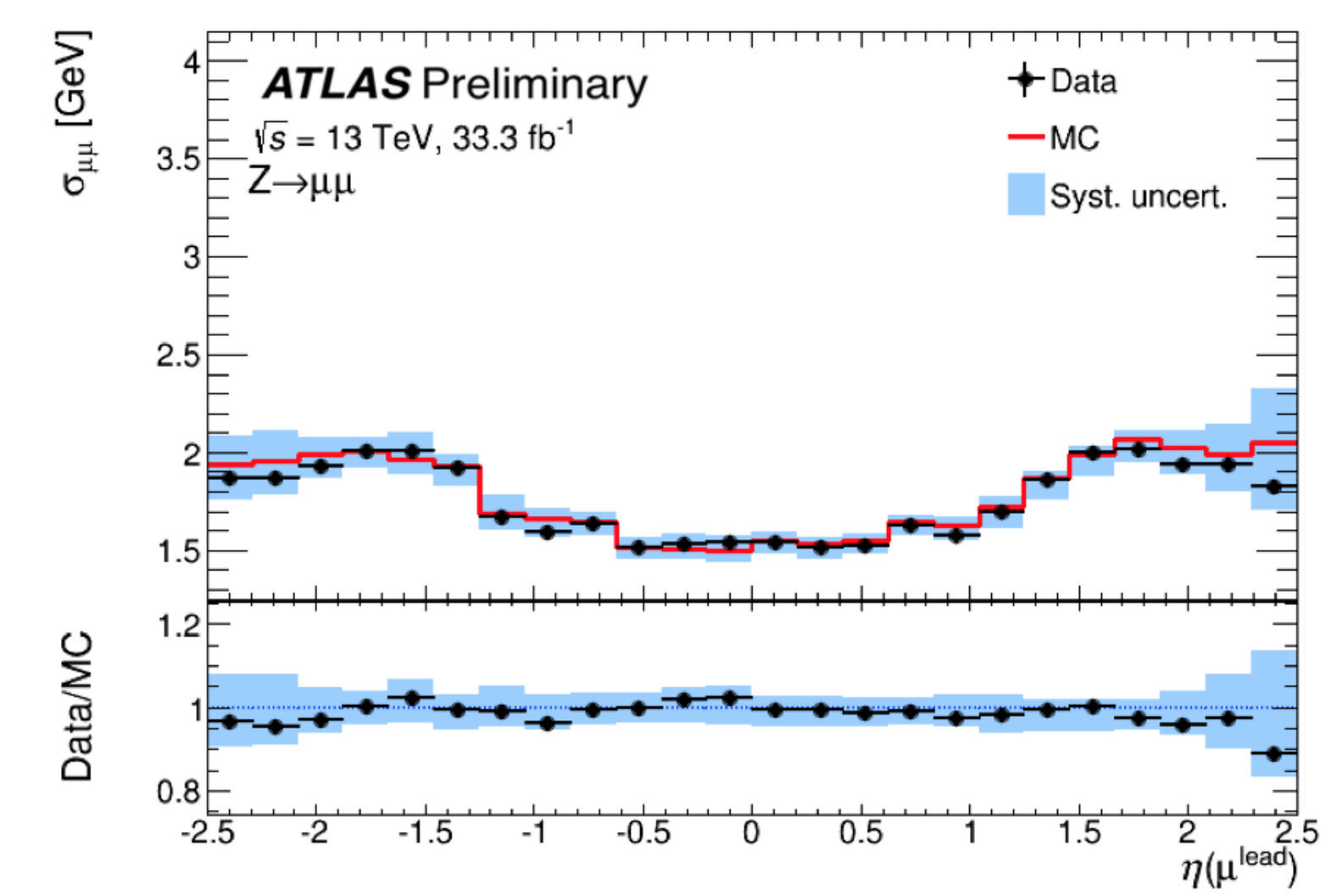}}
	\caption{Muon momentum scale and resolution for combined muons, measured in $Z\rightarrow\mu\mu$ decays. The results are shown in various $\eta$ bins~\cite{ATLAS_muon}.\label{fig:ATLAS_muon_momentum_scale_resol}}
\end{figure}

The momentum scale and resolution are validated by comparing the $Z$ and $J/\psi$ mass spectra between data and corrected MC. The data-to-MC comparison in Fig.~\ref{fig:ATLAS_muon_momentum_scale_resol}~\cite{ATLAS_muon} shows a description at the per-mill level and a resolution at the per-cent level for ATLAS.

\section{Conclusion}

The performance results of the ATLAS and CMS experiments were reported for leptons and photons using Run 2 data (2015 and 2016 included). Good performance was achieved despite the increased pile-up. Strategies vary for the two experiments due to differences in the technologies used, but good results are quite similar.

\end{document}